\documentclass[%
 reprint,
nofootinbib,
 amsmath,amssymb,
 aps,
]{revtex4-1}

\usepackage{graphicx}
\usepackage{dcolumn}
\def\hyp{\mathsf{y}}
\usepackage{bm}
\usepackage[multiple]{footmisc}

\newcommand{\beq}{\begin{equation}}
\newcommand{\eeq}{\end{equation}}
\newcommand{\bea}{\begin{eqnarray}}
\newcommand{\eea}{\end{eqnarray}}
\newcommand{\nn}{\nonumber \\}
\newcommand{\Lagr}{\mathcal{L}}

\makeatletter
\usepackage{tikz}
\newcommand*\circled[2][1.6]{\tikz[baseline=(char.base)]{
    \node[shape=circle, draw, inner sep=1pt,
        minimum height={\f@size*#1},] (char) {\vphantom{WAH1g}#2};}}
\makeatother

\begin{document}

\title{Ward Identities for the Standard Model Effective Field Theory}

\author{Tyler Corbett}
\email{corbett.t.s@gmail.com}
\author{Andreas Helset}
\email{ahelset@nbi.ku.dk}
\author{Michael Trott}
\email{michael.trott@cern.ch}
\affiliation{%
 Niels Bohr International Academy and Discovery Center,
 Niels Bohr Institute,
 University of Copenhagen, Blegdamsvej 17, DK-2100 Copenhagen, Denmark
}%

\date{\today}

\begin{abstract}
We derive Ward identities for the Standard Model Effective Field Theory using the background field method.
The resulting symmetry constraints on the Standard Model Effective Field Theory are basis independent, and constrain the
perturbative and power-counting expansions. 
A geometric description of the field connections, and real representations for
the $\rm SU(2)_L \times U(1)_Y$ generators, underlies the derivation.

\end{abstract}

\maketitle


{\bf Introduction.}
The Standard Model (SM) is an incomplete description of observed phenomena in nature. However, explicit evidence of new 
long-distance propagating states is lacking. Consequently, the SM is usefully thought of as an Effective Field Theory (EFT)
for measurements and data analysis, with characteristic energies proximate to the Electroweak scale
($\sqrt{2 \, \langle H^\dagger H} \rangle \equiv \bar{v}_T$) -- such as those made at the LHC or lower energies.

The Standard Model Effective Field Theory (SMEFT) is based on assuming that physics beyond the SM
is present at scales $\Lambda >\bar{v}_T$. The SMEFT also assumes
that there are no light hidden states in the spectrum with couplings
to the SM; and a $\rm SU(2)_L$ scalar doublet ($H$) with hypercharge
$\hyp_h = 1/2$ is present in the EFT.

A power-counting expansion in
the ratio of scales $\bar{v}_T/\Lambda <1$ defines the SMEFT Lagrangian as
\begin{align}
	\Lagr_{\textrm{SMEFT}} &= \Lagr_{\textrm{SM}} + \Lagr^{(5)}+\Lagr^{(6)} +
	\Lagr^{(7)} + \dots,  \\ \nonumber \Lagr^{(d)} &= \sum_i \frac{C_i^{(d)}}{\Lambda^{d-4}}\mathcal{Q}_i^{(d)}
	\quad \textrm{ for } d>4.
\end{align}
The higher-dimensional operators $\mathcal{Q}_i^{(d)}$ are labelled with a  mass dimension $d$ superscript,
and multiply unknown, dimensionless Wilson coefficients $C_i^{(d)}$. The sum over $i$, after non-redundant operators are removed with field redefinitions
of the SM fields, runs over the operators in a particular operator basis. In this paper we use
the Warsaw basis \cite{Grzadkowski:2010es}. However, the main results are formulated in a basis independent manner
and constrain relationships between Lagragian parameters due to the linear realization of
$\rm SU(2)_L \times U(1)_Y$ in the SMEFT. 

The SMEFT is a powerful practical tool, but it is also a well-defined
field theory. Many formal field-theory issues also have a new representation in the SMEFT. This can lead
to interesting subtleties, particularly when developing SMEFT analyses beyond leading order.
When calculating beyond leading order in the loop ($\hbar$) expansion, renormalization is required.
The counterterms for the SMEFT at dimension five \cite{Babu:1993qv, Antusch:2001ck}, and six \cite{Jenkins:2013zja,Jenkins:2013wua,Alonso:2013hga,Alonso:2014zka}
are known and preserve the $\rm SU(3) \times SU(2) \times U(1)$ symmetry
of the SM. Such unbroken (but non-manifest in some cases) symmetries are also represented in the naive Ward-Takahashi
identities \cite{Ward:1950xp,Takahashi:1957xn} when the
Background Field Method (BFM) \cite{DeWitt:1967ub,tHooft:1973bhk,Abbott:1981ke,Shore:1981mj,Einhorn:1988tc,
Denner:1994xt} is used to gauge fix the theory. In Ref.~\cite{Helset:2018fgq} it was shown how to gauge fix the SMEFT in the BFM in $R_\xi$ gauges, and we use this gauge-fixing procedure
in this work.

The BFM splits the fields in the theory into quantum and classical background fields ($F \rightarrow F + \hat{F}$),
with the latter denoted with a hat superscript. By performing a gauge-fixing procedure that preserves
the background-field gauge invariance, while breaking
explicitly the quantum-field gauge invariance, the Ward identities \cite{Ward:1950xp} are present
in a ``naive manner" -- i.e. the identities are related to those that would be directly inferred from the classical Lagrangian.
This approach is advantageous, as otherwise the gauge-fixing term, and ghost term, of the theory can make
symmetry constraints non-manifest in intermediate steps of calculations.

The BFM gauge-fixing procedure in the SMEFT relies
on a geometric description of the field connections, and real representations for the $\rm SU(2)_L \times U(1)_Y$ generators.
Using this formulation of the SMEFT allows a simple Ward-Takahashi identity to be derived, that constrains the $n$-point
vertex functions. The purpose of this paper is to report this result and derivation.\footnote{Modified Ward identities
in the SMEFT have been discussed in an on-shell scheme in Ref.~\cite{Cullen:2019nnr}.}

{\bf Path integral formulation.}
The BFM generating functional of the SMEFT is given by
\begin{align}
Z[\hat{F},J]=\int \mathcal{D} F \,{\rm det}\left[\frac{\Delta \mathcal{G}^A}{\Delta \alpha^B}\right]e^{i \left(S[F + \hat{F}] + \Lagr_{\textrm{GF}} +
{\rm source \, terms} \right)} \nonumber.
\end{align}
The integration over $d^4x$ is implicit in $\mathcal L_{\rm GF}$.
The generating functional is integrated over the quantum field configurations via $\mathcal{D} F$,
with $F$ field coordinates describing all long-distance propagating states.
$J$ stands for the dependence on the sources which
only couple to the quantum fields \cite{tHooft:1975uxh}. The background fields also effectively act as sources of the quantum fields.
 $S$ is the action, initially classical, and augmented with a renormalization prescription to define loop corrections.

 The scalar Higgs doublet is decomposed into field coordinates $\phi_{1,2,3,4}$, defined with the normalization
 \begin{align}
 	H = \frac{1}{\sqrt{2}}\begin{bmatrix} \phi_2+i\phi_1 \\ \phi_4 - i\phi_3\end{bmatrix}.
 \end{align}
The scalar kinetic term is defined with a field space metric introduced as
 \begin{align}\label{scalarL6}
 	\Lagr_{\textrm{scalar,kin}} = & \frac{1}{2}h_{IJ}(\phi)\left(D_{\mu}\phi\right)^I\left(D^{\mu}\phi\right)^J,
 \end{align}
 where
 $(D^{\mu}\phi)^I = (\partial^{\mu}\delta_J^I - \frac{1}{2}\mathcal{W}^{A,\mu}\tilde\gamma_{A,J}^I)\phi^J$, with real generators ($\tilde\gamma$)
 and structure constants ($\tilde\epsilon^A_{\,\,BC}$) defined in the Appendix.
 The corresponding kinetic term for the $\rm SU(2)_L \times U(1)_Y$ spin-one fields
is 
 \begin{align}\label{WBlagrangian}
 	\Lagr_{\textrm{gauge,kin}} &=-\frac{1}{4}g_{AB}(\phi) \mathcal{W}_{\mu\nu}^A \mathcal{W}^{B,\mu\nu},
 \end{align}
 where $A,B,C, \dots$ run over $\{1,2,3,4\}$, (as do $I,J$)
 and $\mathcal{W}_{\mu\nu}^4=B_{\mu\nu}$. Extending this definition to include the gluons is 
 straight-forward.

 A quantum-field gauge transformation involving these fields is indicated with a $\Delta$, with an infinitesimal quantum gauge parameter $\Delta\alpha^A$.
Explicitly, the transformations are 
\begin{align}
\Delta \mathcal{W}^A_\mu &=  - \tilde{\epsilon}^A_{\, \,BC} \, \Delta \alpha^B \, \left(\hat{\mathcal{W}}^{C, \mu} +\mathcal{W}^{C, \mu} \right) -\partial^\mu (\Delta \alpha^A), \nonumber \\
 \Delta \phi^I &=  - \Delta\alpha^A \, \frac{\tilde\gamma_{A,J}^{I}}{2}\, (\phi^J+ \hat{\phi}^J).
\end{align}
The BFM gauge-fixing term {\it of the quantum fields $\mathcal{W}^{A}$}  is \cite{Helset:2018fgq}
\begin{align}\label{gaugefix}
	\Lagr_{\textrm{GF}} &= -\frac{\hat{g}_{AB}}{2 \, \xi} \mathcal{G}^A \, \mathcal{G}^B, \\
\mathcal{G}^A &\equiv \partial_{\mu} \mathcal{W}^{A,\mu} -
		\tilde\epsilon^{A}_{ \, \,CD}\hat{\mathcal{W}}_{\mu}^C \mathcal{W}^{D,\mu}
    + \frac{\xi}{2}\hat{g}^{AC}
		\phi^{I} \, \hat{h}_{IK} \, \tilde\gamma^{K}_{C,J} \hat{\phi}^J. \nonumber
\end{align}
The introduction of field space metrics in the kinetic terms reflects the geometry of the field space due to the power-counting expansion.
These metrics are the core conceptual difference of the relation between Lagrangian parameters, compared to the SM, in the Ward identities
we derive.
The field spaces defined by these metrics are curved, see Refs.~\cite{Burgess:2010zq,Alonso:2015fsp,Alonso:2016oah}.
The background-field gauge fixing relies on the basis independent transformation properties of
$g_{AB}$ and $h_{IJ}$,\footnote{The explicit forms of $g_{AB}$ and $h_{IJ}$ are basis dependent. The forms of the corrections
for the Warsaw basis at $\mathcal{L}^{(6)}$ are given in Ref.~\cite{Helset:2018fgq}.} and the fields, under background-field gauge transformations ($\delta \hat{F}$)
with infinitesimal local gauge parameters $\delta \hat{\alpha}_A(x)$
given by
\begin{align}\label{backgroundfieldshifts}
\delta \, \hat{\phi}^I &= -\delta \hat{\alpha}^A \, \frac{\tilde{\gamma}_{A,J}^I}{2} \hat{\phi}^J, \nonumber \\
\delta \hat{\mathcal{W}}^{A, \mu} &= - (\partial^\mu \delta^A_B + \tilde{\epsilon}^A_{\, \,BC} \, \, \hat{\mathcal{W}}^{C, \mu}) \delta \hat{\alpha}^B, \nonumber \\
\delta \hat{h}_{IJ} &= \hat{h}_{KJ} \, \frac{\delta \hat{\alpha}^A  \, \tilde{\gamma}_{A,I}^K}{2}+ \hat{h}_{IK} \, \frac{\delta \hat{\alpha}^A  \, \tilde{\gamma}_{A,J}^K}{2}, \nonumber \\
\delta \hat{g}_{AB} &= \hat{g}_{CB} \,\tilde{\epsilon}^C_{\, \,DA} \, \delta \hat{\alpha}^D + \hat{g}_{AC} \,\tilde{\epsilon}^C_{\, \,DB} \, \delta \hat{\alpha}^D, \nonumber \\
\delta \mathcal{G}^X &= -\tilde{\epsilon}^X_{\, \,AB} \, \delta \hat{\alpha}^A \mathcal{G}^B,\nn
\delta f_i &= \Lambda_{A,i}^{j}\,  \hat{\alpha}^A \, f_{j}, \nonumber \\
\delta \bar{f}_i &=   \hat{\alpha}^A \, \bar{f}_j \bar{\Lambda}^{j}_{A,i},
\end{align}
where we have left the form of the transformation of the fermion fields implicit.
Here $i,j$ are flavour indicies.
The background-field gauge invariance  of the generating functional, i.e.
\begin{align}
\frac{\delta Z [\hat{F},J]}{\delta \hat{\alpha}^A} &= 0,
\end{align}
 is established by using these gauge transformations in conjunction with
the linear change of variables on the quantum fields.

The generating functional of connected Green's functions is given by
\begin{align}
W[\hat{F},J] &= - i \log Z[\hat{F},J],
\end{align}
where $J = \{J^A_{\mu}, J^I_{\phi},J_{f},J_{\bar{f}}\}$.
As usual the effective action is the Legendre transform
\begin{align}
\Gamma [\hat{F},\tilde{F}] &= W[\hat{F},J] - \int dx^4 J \cdot \tilde{F} \vert_{\tilde{F} =\frac{\delta W}{\delta J}}.
\end{align}
Here our notation is chosen to match Ref.~\cite{Dekens:2019ept}.
$S$-matrix elements are constructed via \cite{Abbott:1983zw,Denner:1996gb,Dekens:2019ept}
\begin{align}
	\label{eq:gammaFull}
	\Gamma^{\rm full} [\hat{F},0] &= \Gamma [\hat{F},0]+ i \int d^4 x \mathcal{L}_{\textrm{GF}}^{\textrm{BF}}.
\end{align}
The last term in Eq.~\eqref{eq:gammaFull} is a gauge-fixing term for the background fields, formally independent from Eq.~\eqref{gaugefix}, and introduced to define the propagators of the background fields.

Finally, we define a generating functional of connected Green's functions $W_c[\hat{J}]$
as a further Legendre transform \cite{Denner:1996gb}
\begin{align}
W_c[\hat{J}] &= \Gamma^{\rm full} [\hat{F}] + i \int d^4 x \left[\sum_{\hat{F}} \hat{J}_{\hat{F}^\dagger} \hat{F} + \sum_f(\bar{f} \hat{J}_{\bar{f}} + \hat{J}_{f} f) \right].
\end{align}
with $ \hat{F} =\{\mathcal{W}^A, \phi^I\}$ and 
\begin{align}
i  \hat{J}_{\hat{F}^\dagger} &= - \frac{\delta \Gamma^{\rm full}}{\delta \hat{F}}, & i  \hat{J}_{f} &= - \frac{\delta \Gamma^{\rm full}}{\delta \bar{f}},
& i  \hat{J}_{\bar{f}} &=  \frac{\delta \Gamma^{\rm full}}{\delta f}, \nonumber \\
\hat{F} &= \frac{\delta W_c}{i \delta \hat{J}_{\hat{F}^\dagger}},
& f &= \frac{\delta W_c}{i \delta \hat{J}_{\bar{f}}},
& \bar{f} &= - \frac{\delta W_c}{i \delta \hat{J}_{f}}.
\end{align}

{\bf{Weak eigenstate Ward identities.}}
The BFM Ward identities follow from the invariance of $\Gamma [\hat{F},0]$ under background-field gauge transformations,
\begin{align}
\frac{\delta \Gamma [\hat{F},0]}{\delta \hat{\alpha}^B} &= 0.
\end{align}
In position space, the identities  are
\begin{align}
0 =&  \left(\partial^\mu \delta^A_B - \tilde{\epsilon}^A_{\, \,BC} \, \, \hat{\mathcal{W}}^{C, \mu}\right)
\frac{\delta \Gamma}{\delta \hat{\mathcal{W}}_A^{\mu}} - \frac{\tilde{\gamma}_{B,J}^I}{2} \hat{\phi}^J \frac{\delta \Gamma}{\delta \hat{\phi}^I} \nonumber \\
& +\sum_j \left(\bar{f}_j \bar{\Lambda}_{B,i}^{j} \, \frac{\delta \Gamma}{\delta \bar{f}_{i}}
-  \frac{\delta \Gamma}{\delta f_{i}} \Lambda_{B,j}^{i} f_j \right).
\end{align}
For some $n$-point function Ward identities, the background fields are set to their vacuum expectation values.
When this is defined through the minimum of the classical action $S$, where the scalar potential is a function of $H^\dagger H$, which we denote as	$\langle \,\rangle$.
For example, the scalar vev defined in this manner is through $\sqrt{2 \, \langle H^\dagger H} \rangle \equiv \bar{v}_T$
and explicitly $\langle \phi^J \rangle$ with an entry set to the numerical value of the vev does not transform via $\tilde{\gamma}_{A,J}^I$.

A direct relation follows between the tadpoles (i.e. the one point functions $\delta\Gamma/\delta\hat\phi^I$) and, $\langle\hat\phi^J\rangle$, given by
\begin{align}
	0 = \partial^\mu\frac{\delta\Gamma}{\delta\hat{\mathcal{W}}^{B,\mu}} - \frac{\tilde\gamma^I_{B,J}}{2}
	\langle\hat\phi^J\rangle \frac{\delta\Gamma}{\delta\hat\phi^I}.
\end{align}
Requiring a Lorentz-invariant vacuum sets the tadpoles for the gauge fields to zero. Thus, for
the scalars
\begin{align}
	\label{eq:tadpole}
	0 = \frac{\tilde\gamma^I_{B,J}}{2}\langle\hat\phi^J\rangle \frac{\delta\Gamma}{\delta\hat\phi^I}.
\end{align}
$\gamma_{B} \langle \phi^J \rangle \neq 0$
and the unbroken combination $(\gamma_3 + \gamma_4) \langle \phi^J \rangle = 0$ corresponds to $\rm U(1)_{em}$.
Eq.~\eqref{eq:tadpole} with $B=3,4$ does not given linearly independent constraints.
This leads to the requirement of a further renormalization condition to define the tadpole $\delta\Gamma/\delta\hat\phi^4$ to vanish.

The Ward identities for the two-point functions are
\begin{align}
	0 =& \partial^\mu \frac{\delta^2\Gamma}{\delta \hat{\mathcal{W}}^{A,\nu}\delta \hat{\mathcal{W}}^{B,\mu}}
	- \frac{\tilde\gamma^I_{B,J}}{2}\langle\hat\phi^J\rangle
	\frac{\delta^2\Gamma}{\delta \hat{\mathcal{W}}^{A,\nu}\delta\hat\phi^I}, \\
	0 =& \partial^\mu \frac{\delta^2\Gamma}{\delta\hat\phi^K\delta  \hat{\mathcal{W}}^{B,\mu}}
	- \frac{\tilde\gamma^I_{B,J}}{2}\left(\langle\hat\phi^J\rangle
	\frac{\delta^2\Gamma}{\delta\hat\phi^K\delta\hat\phi^I}
	+ \delta^J_K\frac{\delta \Gamma}{\delta\hat\phi^I}\right).
\end{align}
The three-point Ward identities are
\begin{align}
	0 =& \partial^\mu \frac{\delta^3\Gamma}{\delta \overline{f}_k\delta f_l\delta \hat{\mathcal{W}}^{B,\mu}}
	- \frac{\tilde\gamma^I_{B,J}}{2}\langle\hat\phi^J\rangle \frac{\delta^3 \Gamma}{\delta
		\overline{f}_k\delta f_l \delta\hat\phi^I} \nonumber \\
	&+ \overline{\Lambda}^k_{B,i} \frac{\delta^2\Gamma}{\delta\overline{f}_i\delta f_l}
	- \frac{\delta^2\Gamma}{\delta\overline{f}_k\delta f_i} \Lambda^i_{B,l},\\
	0 =& \partial^\mu\frac{\delta^3\Gamma}{\delta \hat W^{A,\nu}\delta\hat W^{B,\mu}
	\delta\hat W^{C,\rho}}
	- \tilde\epsilon^D_{\,\,BC}\frac{\delta^2\Gamma}{\delta\hat{\mathcal{W}}^{D,\rho}
	\delta\hat{\mathcal{W}}^{A,\nu}} \nonumber \\
	&-\frac{\tilde\gamma^I_{B,J}}{2}\langle\hat\phi^J\rangle
	\frac{\delta^3\Gamma}{\delta\hat\phi^I\delta\hat{\mathcal{W}}^{A,\nu}\delta\hat{\mathcal{W}}^{C,\rho}},
  \end{align}
  \begin{align}
	0 =& \partial^\mu \frac{\delta^3\Gamma}{\delta\hat{\mathcal{W}}^{A,\nu}\delta\hat{\mathcal{W}}^{B,\mu}
\delta\hat \phi^K} - \tilde\epsilon^D_{\,\,BA}\frac{\delta^2\Gamma}{\delta\hat{\mathcal{W}}^{D,\nu}
\delta\hat\phi^K}  \nonumber \\
&- \frac{\tilde\gamma^{I}_{B,J}}{2}\left(\langle\hat\phi^J\rangle
\frac{\delta^3\Gamma}{\delta\hat{\mathcal{W}}^{A,\nu}\delta\hat\phi^I\delta\hat\phi^K}
+ \delta^J_K \frac{\delta^2\Gamma}{\delta\hat{\mathcal{W}}^{A,\nu}\delta\hat\phi^I}\right), \\
	0 =& \partial^\mu\frac{\delta^3\Gamma}{\delta\hat{\mathcal{W}}^{B,\mu}\delta\hat\phi^K\delta\hat\phi^L}
	- \frac{\tilde\gamma^I_{B,J}}{2}\langle\hat\phi^J\rangle
	\frac{\delta^3\Gamma}{\delta\hat\phi^I\delta\hat\phi^K\delta\hat\phi^L} \nonumber\\
	&- \frac{\tilde\gamma^I_{B,J}}{2}\left(
	 \delta^J_K\frac{\delta^2\Gamma}{\delta\hat\phi^I\delta\hat\phi^L}
	+ \delta^J_L\frac{\delta^2\Gamma}{\delta\hat\phi^I\delta\hat\phi^K}\right).
\end{align}

{\bf{Mass eigenstate Ward identities.}}
The mass eigenstate SM Ward identities in the BFM are summarized in Ref.~\cite{Denner:1994xt}.
The tranformation of the gauge fields, gauge parameters and scalar fields into mass eigenstates in the SMEFT is
\begin{align}\label{basicrotations}
\hat{\mathcal{W}}^{A,\nu} &=  \sqrt{g}^{AB} U_{BC} \mathcal{\hat{A}}^{C,\nu}, \\
\hat{\alpha}^{A} &= \sqrt{g}^{AB} U_{BC} \mathcal{\hat{\beta}}^{C},\\
\hat{\phi}^{J} &= \sqrt{h}^{JK} V_{KL} \hat{\Phi}^{L},
\end{align}
with $\hat{\mathcal{A}}^C =(\hat{\mathcal{W}}^+,\hat{\mathcal{W}}^-,\hat{\mathcal{Z}},\hat{\mathcal{A}})$,
$\hat{\Phi}^L = \{\hat{\Phi}^+,\hat{\Phi}^-,\hat{\chi},\hat{H}^0 \}$.
This follows directly from the formalism in Ref.~\cite{Helset:2018fgq} (see also Ref.~\cite{Misiak:2018gvl}).
The matrices $U,V$ are unitary, with $\sqrt{g}^{AB}\sqrt{g}_{BC} \equiv \delta^A_C$ and
$  \sqrt{h}^{AB} \sqrt{h}_{BC}\equiv \delta^A_C$. The square root metrics are understood to be matrix square roots
and the entries are $\langle \rangle$ of the field space metrics entries.
The combinations $ \sqrt{g} U$ and $\sqrt{h} V$ perform the mass eigenstate rotation for the vector and scalar fields, and bring the corresponding
kinetic term to canonical form, including higher-dimensional-operator corrections.
We define the mass-eigenstate transformation matrices
\begin{align*}\label{wardrotations}
{\mathcal U}^{A}_C &=  \sqrt{g}^{AB} U_{BC},&  ({\mathcal U^{-1}})^{D}_F &= U^{DE}  \sqrt{g}_{\, EF}  , \\
 {\mathcal V}^{A}_C &= \sqrt{h}^{AB} V_{BC}, &  ({\mathcal V^{-1}})^{D}_F &= V^{DE}  \sqrt{h}_{\, EF} ,
\end{align*}
to avoid a proliferation of index contractions. 
The structure constants and generators, transformed to those corresponding to the mass eigenstates, are defined as
\begin{align*}
{ {\bm \epsilon}}^{C}_{\, \,GY} &= ({\mathcal U^{-1}})^C_A \tilde{\epsilon}^{A}_{\, \,DE} \,  {\mathcal U}^D_G \,
 {\mathcal U}^E_Y, &
 {\bm \gamma}_{G,L}^{I} &= \frac{1}{2}\tilde{\gamma}_{A,L}^{I} \, {\mathcal U}^A_G,\nn
{\bm \Lambda}^i_{X,j} &=\Lambda_{A,j}^{i} \, {\mathcal U}^A_X.
\end{align*}
The background-field gauge transformations in the mass eigenstate are
\begin{align}
\delta \hat{\mathcal{A}}^{C,\mu} &= - \left[\partial^\mu \delta^C_G + { {\bm \epsilon}}^{C}_{\, \,GY} \hat{\mathcal{A}}^{Y,\mu} \right] \delta \hat{\beta}^G, \nn
 \delta \hat{\Phi}^{K} &=- ({\mathcal V^{-1}})^K_I \,{\bm \gamma}_{G,L}^{I} \, {\mathcal V}^L_N \hat{\Phi}^{N} \delta \hat{\beta}^G.
\end{align}
The Ward identities are then expressed compactly as
\bea
0 &=& \frac{\delta \Gamma}{\delta \hat{\beta}^G}
\\
&=& \partial^\mu \frac{\delta \Gamma}{\delta \hat{\mathcal{A}}^{X,\mu}}
+\sum_j \left(\bar{f}_j  \overline{\bm \Lambda}^j_{X,i} \, \frac{\delta \Gamma}{\delta \bar{f}_{i}}
-  \frac{\delta \Gamma}{\delta f_{i}} {\bm \Lambda}^i_{X,j} f_j \right) \nn
&-&
\frac{\delta \Gamma}{\delta \hat{\mathcal{A}}^{C\mu}} {\bm \epsilon}^{C}_{\, \,XY}  \hat{\mathcal{A}}^{Y \mu}
 - \frac{\delta \Gamma}{\delta \hat{\Phi}^K} ({\mathcal V^{-1}})^K_I {\bm \gamma}_{X,L}^{I} {\mathcal V}^L_N \hat{\Phi}^N. \nonumber
\eea
In this manner, the ``naive" form of the Ward identities is maintained.
The BFM Ward identities in the SMEFT take the same
form as those in the SM up to 
terms involving the tadpoles.
This is the case once a consistent redefinition of couplings, masses and fields is made.

{\bf Two-point function Ward Identities.}
The Ward identities for the two-point functions take the form
\bea
0 &=& \partial^\mu \frac{\delta^2 \Gamma}{\delta \hat{\mathcal{A}}^{X \mu} \delta  \hat{\mathcal{A}}^{Y \nu}} -
\frac{\delta^2 \Gamma}{\delta \hat{\mathcal{A}}^{Y \nu} \delta \hat{\Phi}^K} ({\mathcal V^{-1}})^K_I {\bm \gamma}_{X,L}^{I}  {\mathcal V}^L_N \langle \hat{\Phi}^N \rangle, \nn
0 &=& \partial^\mu \frac{\delta^2 \Gamma}{\delta \hat{\mathcal{A}}^{X \mu}  \delta \hat{\Phi}^O}
-
\frac{\delta^2 \Gamma}{\delta \hat{\Phi}^{K} \delta \hat{\Phi}^O} ({\mathcal V^{-1}})^K_I {\bm \gamma}_{X,L}^{I} {\mathcal V}^L_N  \langle \hat{\Phi}^N \rangle \nn
&-&
\frac{\delta \Gamma}{\delta \hat{\Phi}^{K}} ({\mathcal V^{-1}})^K_I {\bm \gamma}_{X,L}^{I} {\mathcal V}^L_O.
\eea

{\bf Photon Identities}
The Ward identities for the two-point functions involving the photon are given by
\begin{align}
0 &= \partial^\mu \frac{\delta^2 \Gamma}{\delta \hat{\mathcal{A}}^{4 \mu} \delta  \hat{\mathcal{A}}^{Y \nu}}, &
0 &= \partial^\mu \frac{\delta^2 \Gamma}{\delta \hat{\mathcal{A}}^{4 \mu} \delta  \hat{\Phi}^{I}}.
\end{align}
Using the convention of Ref.~\cite{Denner:1994xt} for the decomposition of the vertex function
\begin{align}
-i \Gamma^{\hat{V},\hat{V}'}_{\mu \nu}(k,-k)&=
\left(-g_{\mu \nu} k^2 + k_\mu k_\nu + g_{\mu \nu} M_{\hat{V}}^2\right)\delta^{\hat{V} \hat{V}'}, \nonumber \\
&+\left(-g_{\mu \nu} +\frac{k_\mu k_\nu}{k^2}  \right) \Sigma_{T}^{\hat{V},\hat{V}'}- \frac{k_\mu k_\nu}{k^2}
\Sigma_{L}^{\hat{V},\hat{V}'},\nonumber
\end{align}
an overall normalization
factors out of the photon two-point Ward identities compared to the SM, and
\begin{align}
	\Sigma^{\mathcal{\hat{A}},\mathcal{\hat{A}}}_{L,\textrm{SMEFT}}(k^2) &= 0, & \Sigma^{\mathcal{\hat{A}},\mathcal{\hat{A}}}_{T,\textrm{SMEFT}}(0) &= 0.
 \end{align}
The latter result follows from analyticity at $k^2 =0$.

{\bf $\boldsymbol{\mathcal{W}}^\pm, \boldsymbol{\mathcal{Z}}$ Identities.}
Directly, one finds the identities
\bea
0 &=& \partial^\mu \frac{\delta^2 \Gamma}{\delta \hat{\mathcal{A}}^{3 \mu} \delta  \hat{\mathcal{A}}^{Y \nu}} -
\bar{M}_Z \,  \frac{\delta^2 \Gamma}{\delta  \hat{\Phi}^{3} \delta \hat{\mathcal{A}}^{Y \nu}}, \\
0 &=& \partial^\mu \! \! \frac{\delta^2 \Gamma}{\delta \hat{\mathcal{A}}^{3 \mu} \delta  \hat{\Phi}^{I} }
-\bar{M}_Z  \frac{\delta^2 \Gamma}{\delta  \hat{\Phi}^{3} \delta  \hat{\Phi}^{I} }  \\
&+& \frac{\bar{g}_Z}{2} \frac{\delta \Gamma}{\delta  \hat{\Phi}^{4}}  \left(\sqrt{h}_{[4,4]}  \sqrt{h}^{[3,3]} - \sqrt{h}_{[4,3]} \sqrt{h}^{[4,3]}\right) \delta^3_I  \nonumber \\
&-& \frac{\bar{g}_Z}{2} \frac{\delta \Gamma}{\delta  \hat{\Phi}^{4}}  \left(\sqrt{h}_{[4,4]}  \sqrt{h}^{[3,4]} - \sqrt{h}_{[4,3]} \sqrt{h}^{[4,4]}\right) \delta^4_I, \nonumber
\eea
and
\bea
0 &=& \partial^\mu \frac{\delta^2 \Gamma}{\delta \hat{\mathcal{W}}^{\pm \mu} \delta  \hat{\mathcal{A}}^{Y \nu}} \pm
i \bar{M}_W \frac{\delta^2 \Gamma}{\delta  \hat{\Phi}^{\pm} \delta \hat{\mathcal{A}}^{Y \nu}},  \\
0 &=& \partial^\mu \frac{\delta^2 \Gamma}{\delta \hat{\mathcal{W}}^{\pm \mu} \delta  \hat{\Phi}^{I}}
\pm i \bar{M}_W \frac{\delta^2 \Gamma}{\delta  \hat{\Phi}^{\pm} \delta \hat{\Phi}^{I}} \\
&\mp& \frac{i \bar{g}_2}{4} \frac{\delta \Gamma}{\delta  \hat{\Phi}^{4}}
\left(\sqrt{h}_{[4,4]}\mp i \sqrt{h}_{[4,3]} \right) \times \nn
&\,&\left[(\sqrt{h}^{[1,1]}+ \sqrt{h}^{[2,2]} \mp i \sqrt{h}^{[1,2]} \pm i \sqrt{h}^{[2,1]}) \delta^{\mp}_I \right.\nn
&-& \left.(\sqrt{h}^{[1,1]}- \sqrt{h}^{[2,2]} \pm i \sqrt{h}^{[1,2]} \pm i \sqrt{h}^{[2,1]}) \delta^{\pm}_I\right]. \nonumber
\eea
These identities have the same structure as in the SM. The main differences are the factors multiplying the tadpole terms.
By definition, the vev is defined as
$\sqrt{2 \, \langle H^\dagger H} \rangle \equiv \bar{v}_T$. The substitution of
the vev leading to the $\hat{\mathcal{Z}}$ boson mass in the SMEFT ($\bar{M}_Z$)
absorbs a factor in the scalar mass-eigenstate transformation matrix as
$\sqrt{2 \, \langle H^\dagger H} \rangle = \sqrt{2 \, \langle H^\dagger {\mathcal V^{-1}}{\mathcal V} H \rangle}$.
If a scheme is chosen so that $\delta\Gamma/\delta\hat\phi^4$ vanishes, then rotation to the mass eigenstate basis of the one-point vector
$\delta\Gamma/\delta\hat\phi^i$ are still vanishing in each equation above. 
One way to tackle tadpole corrections is to use the
FJ tadpole scheme, for discussion see Ref.~\cite{Fleischer:1980ub,Denner:2018opp}.

{\bf $\boldsymbol{\mathcal{A}},\boldsymbol{\mathcal{Z}}$ Identities.}
The mapping of the SM Ward identites for $\Gamma_{AZ}$ in the background field method given in Ref.~\cite{Denner:1994xt}
to the SMEFT is
\begin{align}
0 = \partial^\mu \frac{\delta^2\Gamma}{\delta \mathcal{\hat{A}}^{\nu}\delta \hat{\mathcal{Z}}^{\mu}}.
\end{align}
As an alternative derivation, the mapping between the mass eigenstate $(Z,A)$ fields in the
SM and the SMEFT ($\mathcal{Z},\mathcal{A}$) reported in Ref.~\cite{Brivio:2017btx} directly follows
from Eq.~\eqref{wardrotations}. Input parameter scheme dependence drops out when considering the two-point function
$\Gamma_{AZ}$ in the SM mapped to the SMEFT and a different overall normalization factors out.
One still finds $\Sigma^{\mathcal{\hat{A}},\mathcal{\hat{Z}}}_{L,{\rm SMEFT}}(k^2) = 0$ and, as a consequence of analyticity at $k^2 =0$,
$\Sigma^{\mathcal{\hat{A}},\mathcal{\hat{Z}}}_{T,{\rm SMEFT}}(0) = 0$. This result has been used in the BFM calculation
reported in Ref.~\cite{Hartmann:2015oia,Hartmann:2015aia}.

{\bf Conclusions.}
We have derived Ward identities for the SMEFT, 
constraining both the perturbative and power-counting expansions. The results presented already provide a clarifying explanation to
some aspects of the structure of the SMEFT that has been determined at tree level. The utility of these results is expected
to become clear as studies of the SMEFT advance to include sub-leading corrections.

\begin{acknowledgments}
We acknowledge support from the Carlsberg Foundation, the Villum Fonden and the Danish National Research Foundation (DNRF91)
  through the Discovery center.
We thank W. Dekens, A. Manohar, G. Passarino and P. Stoffer for discussions and/or comments on the draft, and P. van Nieuwenhuizen for his AQFT notes.
\end{acknowledgments}
\newpage
{\bf Notation.}
The metric forms and rotations to $\mathcal{L}^{(6)}$ in the Warsaw basis are explicitly
\cite{Grinstein:1991cd,Alonso:2013hga}
\begin{align}
 \sqrt{g}^{AB} &= \begin{bmatrix}
	1+\tilde{C}_{HW} & 0 & 0 & 0 \\
	0 & 1+\tilde{C}_{HW} & 0 & 0 \\
	0 & 0 & 1+\tilde{C}_{HW} & -\frac{\tilde{C}_{HWB}}{2} \\
	0 & 0 & -\frac{\tilde{C}_{HWB}}{2} & 1+\tilde{C}_{HB}
	\end{bmatrix}, \nonumber \\
	U_{BC} &= \begin{bmatrix}
		\frac{1}{\sqrt{2}} & \frac{1}{\sqrt{2}} & 0 & 0 \\
		\frac{i}{\sqrt{2}} & \frac{-i}{\sqrt{2}} & 0 & 0 \\
		0 & 0 & c_{\overline{\theta}} & s_{\overline{\theta}} \\
		0 & 0 & -s_{\overline{\theta}} & c_{\overline{\theta}}
	\end{bmatrix}, \nonumber \\
\sqrt{h}^{IJ} &= \begin{bmatrix}
	1 & 0 & 0 & 0 \\
	0 & 1 & 0 & 0 \\
	0 & 0 & 1-\frac{1}{4}\tilde{C}_{HD} & 0 \\
	0 & 0 & 0 & 1+\tilde{C}_{H\Box}-\frac{1}{4}\tilde{C}_{HD}
	\end{bmatrix}, \nonumber \\
	V_{JK} &= \begin{bmatrix}
		\frac{-i}{\sqrt{2}} & \frac{i}{\sqrt{2}} & 0 & 0 \\
		\frac{1}{\sqrt{2}} & \frac{1}{\sqrt{2}} & 0 & 0 \\
	0 & 0 & -1 & 0 \\
	0 & 0 & 0 & 1
	\end{bmatrix}.
\end{align}
The notation for dimensionless Wilson coefficients is $\tilde{C}_i = \bar{v}_T^2 C_i/\Lambda^2$.
The convention for
$s_{\bar{\theta}}$ here has a sign consistent with Ref.~\cite{Alonso:2013hga}, which has an
opposite sign compared to Ref.~\cite{Denner:1994xt}. For details and explicit
results on couplings for the SMEFT including $\mathcal{L}^{(6)}$ corrections in the Warsaw basis,
we note that we are consistent in notational conventions with Ref.~\cite{Alonso:2013hga}.

The generators are given as
\begin{align}
	\gamma_{1,J}^{I} &=  \begin{bmatrix}
		0 & 0 & 0 & -1 \\
		0 & 0 & -1 & 0 \\
		0 & 1 & 0 & 0 \\
		1 & 0 & 0 & 0
	\end{bmatrix}, &
	\gamma_{2,J}^{I} &= \begin{bmatrix}
		0 & 0 & 1 & 0 \\
		0 & 0 & 0 & -1 \\
		-1 & 0 & 0 & 0 \\
		0 & 1 & 0 & 0
	\end{bmatrix}, \nonumber \\
	\gamma_{3,J}^{I} &=  \begin{bmatrix}
		0 & -1 & 0 & 0 \\
		1 & 0 & 0 & 0 \\
		0 & 0 & 0 & -1 \\
		0 & 0 & 1 & 0
	\end{bmatrix}, &
	\gamma_{4,J}^{I} &= \begin{bmatrix}
		0 & -1 & 0 & 0 \\
		1 & 0 & 0 & 0 \\
		0 & 0 & 0 & 1 \\
		0 & 0 & -1 & 0
	\end{bmatrix}.
\end{align}
The  $\gamma_{4}$ generator is used for the $\rm U(1)_Y$ embedding.
The couplings are absorbed into the structure constants and generators leading to tilde superscripts, 
\begin{align}
	\tilde{\epsilon}^{A}_{\, \,BC} &= g_2 \, \epsilon^{A}_{\, \, BC}, \text{ \, \, with } \tilde{\epsilon}^{1}_{\, \, 23} = +g_2,  \nonumber \\
	\tilde{\gamma}_{A,J}^{I} &= \begin{cases} g_2 \, \gamma^{I}_{A,J}, & \text{for } A=1,2,3 \\
		g_1\gamma^{I}_{A,J}, & \text{for } A=4.
					\end{cases}
\end{align}
In mass eigenstate basis, the transformed generators are
\begin{align}
{\bm \gamma}_{1,J}^{I}  &= \frac{\overline{g}_2}{2\sqrt{2}}\begin{bmatrix}
	0 & 0 & i & -1 \\
	0 & 0 & -1 & -i \\
	-i & 1 & 0 & 0 \\
	1 & i & 0 & 0
	\end{bmatrix}, \nonumber \\
	{\bm \gamma}_{2,J}^{I}&= \frac{\overline{g}_2}{2\sqrt{2}}\begin{bmatrix}
	0 & 0 & -i & -1 \\
	0 & 0 & -1 & i \\
	i & 1 & 0 & 0 \\
	1 & -i & 0 & 0
	\end{bmatrix}, \nonumber
\end{align}
\begin{align}
{\bm \gamma}_{3,J}^{I} &= \frac{\overline{g}_Z}{2}\begin{bmatrix}
		0 & -(c_{\overline{\theta}}^2 - s_{\overline{\theta}}^2) & 0 & 0 \\
	(c_{\overline{\theta}}^2 - s_{\overline{\theta}}^2)  & 0 & 0 & 0 \\
	0 & 0 & 0 & -1 \\
	0 & 0 & 1 & 0
	\end{bmatrix}, \nonumber \\
{\bm \gamma}_{4,J}^{I} &= \overline{e}\begin{bmatrix}
	0 & -1 & 0 & 0 \\
	1 & 0 & 0 & 0 \\
	0 & 0 & 0 & 0 \\
	0 & 0 & 0 & 0
	\end{bmatrix}.
\end{align}

{\bf Connected Green's functions formulation}
An alternative approach is to derive the Ward identities in terms of the generating functional for connected Green's functions -- $W_c$.
The non-invariance of $\mathcal{L}_{\textrm{GF}}^{\textrm{BF}}$ under background-field gauge transformations leads to
\begin{align}
\frac{\delta W_c}{\delta \alpha^B} = i \int d^4 x \, \frac{\delta}{\delta\hat\alpha^B}\Lagr_{\textrm{GF}}^{\textrm{BF}}.
\end{align}
We choose the gauge-fixing term {\it for the background fields}
\begin{align}
	\Lagr_{\textrm{GF}}^{\textrm{BF}} &= -\frac{1}{2\xi} \langle{g}_{AB}\rangle {G}^A {G}^B, \\
	G^X &= \partial_\mu \hat W^{X,\mu} + \frac{\xi}{2}\langle {g}^{XC}\rangle (\hat{\phi}^I - \langle \hat \phi^I
	\rangle ) \langle{h}_{IK}\rangle\tilde\gamma^K_{C,J}\langle{\hat{\phi}}^J\rangle. \nonumber
\end{align}
The variation of the gauge-fixing term with respect to the background-gauge parameter is
\begin{align}
&	\frac{\delta}{\delta\hat\alpha^B}\Lagr_{\textrm{GF}}^{\textrm{BF}} = \frac{1}{\xi}\langle{g}_{AD}\rangle
	\left(
	\Box\delta^A_B + i\partial^{\mu} \tilde\epsilon^A_{\,\,BC}\frac{\delta W_c}{\delta J_{\hat W^{C,\mu}}}	\right. \\
&	\left. + \frac{\xi}{2}\langle{g}^{AE}\rangle \frac{\tilde\gamma^I_{B,J}}{2}
	\left( -i\frac{\delta W_c}{\delta J_{\hat\phi^J}} \right)
	\langle{h}_{IK}\rangle
	\tilde\gamma^{K}_{E,L}\langle{\phi}^L\rangle
	\right)
	G^D_{\mathcal{J}}, \nonumber
\end{align}
where
\begin{align}
	G^D_{\mathcal{J}} = -i\partial^\nu \frac{\delta W_c}{\delta J_{\hat W^{D,\nu}}}
	-i \frac{\xi}{2}\langle{g}^{DX}
  \rangle \frac{\delta W_c}{\delta J_{\hat\phi^I}}\langle{h}_{IK}\rangle
	\tilde\gamma^{K}_{X,J}\langle{\phi}^J\rangle. \nonumber
\end{align}

Consider the difference between the vev defined by $\langle \,\rangle$
and an alternate vev denoted by $\langle \phi^J\rangle^\prime$ where the minimum of the action still dictates the
numerical value, but in addition $\langle \phi^J\rangle^\prime$ transforms as $\delta \langle \phi^I\rangle^\prime = \tilde{\gamma}_{A,J}^I \langle \phi^J\rangle^\prime \, \hat{\alpha}^A$.
Replacing all instances of $\langle \rangle$ in the above equations with this expectation value, and related transformation properties on the modified metrics, one finds
\begin{align}
\frac{\delta}{\delta\hat\alpha^B}{\Lagr}_{\textrm{GF}}^{\textrm{BF}} = \frac{1}{\xi}\langle{g}_{BD}\rangle^\prime
	\Box G^D_{\mathcal{J}}.
\end{align}
The two results coincide for on-shell observables, for further discussion this point, and tadpole schemes, see Ref.~\cite{Denner:1996gb}.
We postpone a detailed
discussion of these two approaches to a future publication.\\

\bibliography{bibliography2}

\end{document}